\documentclass[proceedings]{JHEP3}
\usepackage{amsfonts}
\usepackage{amsmath}
\usepackage{epsfig,multicol}

\newbox\mybox

\newcommand\fverb{\setbox\mybox=\hbox\bgroup\verb}
\newcommand\fverbdo{\egroup\medskip\noindent\fbox{\unhbox\mybox}\ }
\newcommand\fverbit{\egroup\item[\fbox{\unhbox\mybox}]}
\conference{On vacuum energies and renomalizability in integrable QFT}
\abstract{We compute for various perturbed conformal field theories the vacuum energies 
by means of the thermodynamic Bethe ansatz. Depending on the infrared and ultraviolet
divergencies of the models, governed by the scaling dimensions of the underlying
perturbed conformal field theory in the ultraviolet, the vacuum energies exhibit
different types of characteristics. In particular, for the homogeneous sine-Gordon
models we observe that once the conformal dimension
of the perturbing scalar field is smaller or greater than 1/2, the vacuum energies
are positive or negative, respectively. This behaviour indicates the need for 
additional ultraviolet counterterms in the latter case. At the transition points
we obtain an infinite vacuum energy, which is partly explainable with the presence of 
several free Fermions in the models studied.}

\title{On vacuum energies and renormalizability in integrable quantum field
theories}
\author{Olalla Castro-Alvaredo$^\circ$ and Andreas Fring$^\bullet$ \\
$^\circ$ Laboratoire de Physique, Ecole Normale Sup\'{e}rieure de Lyon, \\
$\;$ UMR 5672 du CNRS, 46 All\'{e}e d'Italie, 69364 Lyon CEDEX, France\\
$\;$ E-mail: \email{ocastroa@ens-lyon.fr}\\
$^\bullet$  Centre for Mathematical Science, City University, \\
$\;$ Northampton Square, London EC1V 0HB, UK\\
$\;$ E-mail: \email{A.Fring@city.ac.uk}}

\input{tcilatex}

\begin{document}

\section{Introduction}

According to the ideas developed first in \cite{Zamolodchikov:1989fp} a
large class of massive quantum field theories in 1+1 space-time dimensions
can be viewed as perturbed conformal field theories (CFT) with Euclidian
action 
\begin{equation}
\mathcal{S=S}_{\text{CFT}}+\lambda \int d^{2}x\varphi (x)~.  \label{H}
\end{equation}%
Here $\mathcal{S}_{\text{CFT}}$ denotes a fixed point action, $\varphi (x)$
a scalar field with (left, right) conformal dimension $(\Delta ,\Delta )$
and $\lambda $ a coupling constant, scaling with $(1-\Delta ,1-\Delta )$.
The great virtue of such theories is that very often they are integrable and
can be solved exactly, that is to all orders in perturbation theory. Since
the original formulation various non-perturbative techniques have been
developed to study such theories with great success. Nonetheless, once the
CFT is well investigated one may also employ standard perturbative arguments
and unravel the meaning of certain types of behaviour in that more
traditional language. Accordingly, the vacuum expectation value of any local
operator $\mathcal{O}$ can then be computed as%
\begin{equation}
\left\langle \mathcal{O}(z,\bar{z})\right\rangle
=Z^{-1}\sum\limits_{n=0}^{\infty }\frac{(-\lambda )^{n}}{n!}\int
d^{2}z_{1}\ldots d^{2}z_{n}\left\langle \mathcal{O}(z,\bar{z})\varphi (z_{1},%
\bar{z}_{1})\ldots \varphi (z_{n},\bar{z}_{n})\right\rangle _{\text{CFT}}~.
\label{O}
\end{equation}%
Here the normalization factor is in general $Z=\left\langle \exp -\lambda
\int d^{2}z\varphi (z,\bar{z})\right\rangle _{\text{CFT}}$, with $\left.
{}\right\rangle _{\text{CFT}}$ denoting the vacuum state related to $%
\mathcal{S}_{\text{CFT}}$. In quantum field theories such expressions are
plagued by various types of divergencies. First of all one encounters the
infinities due to the self-contraction of the fields, which can be
regularized fairly easily by a normal ordering prescription. Second, one
might have ultraviolet (UV) singularities for $(z-z_{i})\rightarrow 0$. Here
the case $\mathcal{O=}\varphi $ will be important, for which we can
approximate with the help of standard CFT operator product expansion the
integrals in (\ref{O}) as $\sim $ $\int dz_{i}\left\vert z-z_{i}\right\vert
^{-2\Delta }$. Thus for $\Delta <1/2$ the integrals in (\ref{O}) remain
finite, whereas for $\Delta >1/2$ we require in general counterterms to
elliminate the divergencies. Third, one might have infrared (IR)
singularties for $(z-z_{i})\rightarrow \infty $. In the infinite plane it is
usually an intricate issue to handle them \cite{FD1,FD2}. However, when
formulating the theory from the very beginning on a cylinder instead of an
infinite plane the integrals in (\ref{O}) will automatically be IR finite
for $\Delta >0$, as the cylinder radius $R$ constitutes a natural cut off.
The fourth singularity occurring is related to the fact, that even when the
individual integrals in\ (\ref{O}) are finite the entire series will in
general be IR divergent for large $R$.

Supposing now that one is able to compute (\ref{O}) exactly, that is to all
orders in perturbation theory, the different types of renormalization
quantities should be tractable in that context. In fact, the thermodynamic
Bethe ansatz (TBA) \cite{Zamolodchikov:1990cf} is a method which allows such
identifications when $\mathcal{O}$ is taken to be the energy operator. The
above mentioned arguments hold when recalling \cite{Cardy:1988tj} that this
operator is proportional to the perturbing field $\varphi $. Defining then
for the ground state energy $E(R)$ the scaling function $c(R)=-6RE(R)/\pi $
one encounters several types of general behaviours, which can all be brought
into the generic form 
\begin{equation}
c(r)=c_{\text{eff}}+\mathcal{E}_{0}r^{2}+\mathcal{E}_{0}^{^{\prime
}}r^{2}\ln r+\sum\limits_{n=1}^{\infty }\mathcal{E}_{n}\lambda ^{n}f_{n}(r)~.
\label{gb}
\end{equation}%
Usually one uses the dimensionless parameter $r=R/m$ with $m$ being a mass
scale and $\mathcal{E}_{0}$, $\mathcal{E}_{0}^{^{\prime }}$ being finite
real numbers. The function $c(r)$ is normalized in such a way that $c(r=0)$
coincides with the effective central charge $c_{\text{eff}}=c-24\Delta _{%
\text{min}}$, with $c$ being the Virasoro central charge of the underlying
ultraviolet conformal field theory and $\Delta _{\text{min}}$ the smallest
conformal scaling dimension in the model. This constant $c$ has the well
known interpretation as the Casimir energy, which is the vacuum energy on
the cylinder and becomes zero when mapped onto the plane. Viewing (\ref{O})
as resulting from a partition function, the term $\mathcal{E}_{0}r^{2}$ has
to be present in (\ref{gb}), since thermodynamics dictates that for large $r$
the energy has to be proportional to the volume. In quantum field theoretic
terms both $\mathcal{E}_{0}r^{2}$ and $\mathcal{E}_{0}^{^{\prime }}r^{2}\ln r
$ are related to renormalization issues, characterized by the conformal
dimension $\Delta $ as described above. These terms are also needed in order
to ensure that $\lim_{r\rightarrow \infty }c(r)=0$, which one expects for a
purely massive model. Finally, the $f_{n}(r)$ result from the integrals in (%
\ref{O}) and takes on various general forms depending on the regime in which 
$\Delta $ is valued.

In this paper we will discuss more concretely the precise nature of the
expansion (\ref{gb}). We will first recall in section 2 how the TBA can be
used to compute the vacuum energies and in the following sections we discuss
the different regimes for different types of concrete theories, the
homogeneous sine-Gordon (HSG) models \cite%
{Park:1994bx,Fernandez-Pousa:1997hi} and affine Toda field theories (ATFT) 
\cite{Toda,Toda2}. These theories probe several regimes for $\Delta $ and
exhibit different types of behaviours. In particular for the HSG-models,
which are defined in the entire regime $0<\Delta <1$, our results will be
new. Our conclusions are stated in section 6.

\section{Vacuum energies from the TBA}

Let us briefly recall the main steps of how vacuum energies may be computed 
\cite{Zamolodchikov:1990cf} (more details on the arguments can also be found
in \cite{Klassen:1991dx}) non-perturbatively\ with the help of the TBA. One
considers a relativistic theory in which the scattering matrices $%
S_{ij}(\theta )$ for the particles of the type $i$,$j$ with masses $%
m_{i},m_{j}$ are known as functions of the rapidity $\theta $. Then the
entire TBA analysis can be formulated with only two inputs: first\ the
dynamical interaction, which enters via the logarithmic derivative of the
S-matrix $\varphi _{ij}(\theta )=-id\ln S_{ij}(\theta )/d\theta $ and an
assumption on the statistical interaction, which we choose here to be of
fermionic type. The thermodynamic Bethe ansatz equations are then a set of
coupled non-linear integral equations 
\begin{equation}
rm_{i}\cosh \theta =\varepsilon _{i}(\theta ,r)+\sum\limits_{j}[\varphi
_{ij}\ast \ln (1+e^{-\varepsilon _{j}})](\theta ,r)\,,  \tag{TBA}
\label{TBA}
\end{equation}%
where the pseudo-energies $\varepsilon _{i}(\theta ,r)$ are the unknown
quantities. We denote as usual the convolution of two functions by $\left(
f\ast g\right) (\theta )$ $:=1/(2\pi )\int d\theta ^{\prime }f(\theta
-\theta ^{\prime })g(\theta ^{\prime })$. The scaling parameter is related
to the inverse temperature $T$ as $r=m/T$, $m_{l}\rightarrow m_{l}/m$, with $%
m$ being an overall mass scale.\ In \cite{Zamolodchikov:1990cf} it was shown
that when taking the sum and difference of the derivatives $d/dr$(\ref{TBA})
and $d/d\theta $(\ref{TBA})$/r$ one may derive a set of coupled linear
integral equations for the quantities%
\begin{equation}
\psi _{\pm }^{i}(\theta ,r)=\frac{\partial \varepsilon _{i}(\theta ,r)}{%
\partial r}\pm \frac{1}{r}\frac{\partial \varepsilon _{i}(\theta ,r)}{%
\partial \theta },
\end{equation}%
respectively, namely%
\begin{equation}
\psi _{\pm }^{i}(\theta ,r)=m_{i}~e^{\pm \theta }+\sum\limits_{j}[\varphi
_{ij}\ast \frac{1}{e^{\varepsilon _{j}}\pm 1}\psi _{\pm }^{j}](\theta ,r)~.
\label{psi}
\end{equation}%
The strategie is now to solve first the equations (\ref{TBA}) for $%
\varepsilon _{i}(\theta ,r)$ and thereafter (\ref{psi}) for $\psi _{\pm
}^{i}(\theta ,r)$. Once one has carried out the first step, one can already
compute the scaling function 
\begin{equation}
c(r)=\frac{3\,r}{\pi ^{2}}\sum_{i}m_{i}\int\limits_{-\infty }^{\infty
}d\theta \,\cosh \theta \,L_{i}(\theta ,r)\,,  \label{scale}
\end{equation}%
with $L_{i}(\theta ,r)=\ln (1+e^{-\varepsilon _{i}(\theta ,r)})$. Concerning
the status of analytical solutions for (\ref{psi}), it is similar as for the
TBA itself, that is only for free theories \cite{Klassen:1991dx} a closed
solution was found and for interacting theories (\ref{psi}) was only solved
in the extreme ultraviolet limit. Numerical solutions exist even less. Once
it is solved, one may compute the vacuum expectation value of the trace of
the energy momentum tensor, i.e.~vacuum energies%
\begin{eqnarray}
\left\langle T_{\,\,\,\,\mu }^{\mu }\right\rangle &=&-\frac{\pi ^{2}}{3r}%
\frac{d}{dr}c(r)=\frac{1}{2}\sum_{i}m_{i}\left( T_{+}^{i}+T_{-}^{i}\right)
\label{T1} \\
&=&\frac{1}{2}\sum_{i}m_{i}\int\limits_{-\infty }^{\infty }d\theta \,\frac{1%
}{1+e^{\varepsilon _{i}(\theta ,r)}}\left[ \psi _{+}^{i}(\theta
,r)e^{-\theta }+\psi _{-}^{i}(\theta ,r)e^{\theta }\right] ~.
\end{eqnarray}%
In a parity invariant theory we have $\varepsilon _{i}(\theta
,r)=\varepsilon _{i}(-\theta ,r)$ and consequently $\psi _{+}^{i}(\theta
,r)=\psi _{-}^{i}(\theta ,r)$, $T_{+}^{i}=T_{-}^{i}=T^{i}$ such that matters
simplify. We like to keep the treatment here generic for a while as we will
also consider below the homogeneous sine-Gordon models, which are not parity
invariant.

There exists no systematic way to solve the equations (\ref{TBA}) and (\ref%
{psi}) analytically, albeit, numerically this is a solvable problem.
Nonetheless, it is well known that at the fixed points approximations can be
made, such that one can solve (\ref{TBA}) analytically and hence also obtain
analytic expressions for (\ref{scale}) at these points ($r=0$ is one of
them). Likewise we expect to be able to solve (\ref{psi}) for these values
and compute $\left\langle T_{\,\,\,\,\mu }^{\mu }\right\rangle $
analytically. Following now essentially the argumentation of \cite%
{Zamolodchikov:1990cf,Klassen:1991dx}, we need to make only three
assumptions:

\begin{description}
\item[i)] \quad The logarithmic derivative of the scattering matrix in (\ref%
{TBA}) admits an expansion of the form 
\begin{equation}
\varphi _{ij}(\theta )=-\sum\limits_{s}\varphi _{ij}^{(s)}e^{-s\left\vert
\theta \right\vert }~.  \label{A1}
\end{equation}

\item[ii)] \ \ For the first coefficient in (\ref{A1}) we presume
proportionality to the masses 
\begin{equation}
\varphi _{ij}^{(1)}=\rho _{ij}m_{i}m_{j}  \label{A3}
\end{equation}%
for some function $\rho _{ij}$ specific to the particular theory.

\item[iii)] \thinspace \ \thinspace One assumes that 
\begin{equation}
\hat{\varepsilon}_{i}(\theta )-\varepsilon _{i}\ll e^{\theta }\qquad \text{%
for\quad }\theta \ll 0  \label{A2}
\end{equation}%
where the $\varepsilon _{i}$ are the pseudo-energies of the constant TBA
equation and the $\hat{\varepsilon}_{i}(\theta )$ are quantities in the
r-independent TBA-equation 
\begin{equation}
\varphi _{ij}\ast \hat{L}_{j}(\theta )=-\hat{\varepsilon}_{i}(\theta
)+m_{i}e^{\theta }  \label{e1}
\end{equation}%
obtained from (\ref{TBA}) by the shift $\theta \rightarrow \theta +\ln (r/2)$%
, $\varepsilon _{i}(\theta ,r)\rightarrow \hat{\varepsilon}_{i}(\theta )$.
This assumption is usually difficult to justify a priori, but is sustained
in hindsight by meaningful results or supported by numerical data.
\end{description}

\noindent For $\theta \rightarrow -\infty $ one can now derive with (\ref{A2}%
) the equation 
\begin{equation}
\varphi _{ij}\ast \hat{L}_{j}(\theta )=-\varepsilon _{i}+\frac{1}{2\pi }%
e^{\theta }\varphi _{ij}^{(1)}T_{+}^{j}+\mathcal{O}(e^{\eta \theta })
\label{e2}
\end{equation}%
where $\eta \geq 2$. Comparing then (\ref{e1}) and (\ref{e2}) for the parity
invariant case, it follows directly with (\ref{A2}) that%
\begin{equation}
m_{i}=\frac{1}{2\pi }\varphi _{ij}^{(1)}T^{j}~.
\end{equation}%
Finally we deduce the expression for the vacuum expectation value for the
energy momentum tensor with (\ref{A2}) and (\ref{T1}) to%
\begin{equation}
\left\langle T_{\,\,\,\,\mu }^{\mu }\right\rangle =2\pi \sum_{i,j}\rho
_{ij}^{-1}~.  \label{T}
\end{equation}%
This quantity is of course sensitive to above mentioned renormalization
issues and possibly exhibits the distinction between the different regimes
quoted. Furthermore, one has the possibility of comparison, as there are
various other methods to obtain the vacuum energies, such as the truncated
conformal space approach \cite{Yurov:1990yu} or a matching between the
high-energy behaviour of the scattering matrix with a Feynman diagramatic
analysis \cite{Destri:1991ps}.

Let us briefly comment on the different regimes:

\noindent \underline{$\mathbf{0<\Delta <1/2:}$} As mentioned in the
introduction, in this regime the individual integrals in the expansion (\ref%
{O}) are UV and IR convergent term by term when formulated on the cylinder.
From general arguments one finds for the behaviour in (\ref{gb}) that $%
\mathcal{E}_{0}^{^{\prime }}=0$ and $f_{n}(r)=r^{2n(1-\Delta )}$ \cite%
{ADETBA}. From (\ref{gb}) and (\ref{T1}) follows also that we can identify $%
\left\langle T_{\,\,\,\,\mu }^{\mu }\right\rangle |_{r=0}=-\pi ^{2}/3%
\mathcal{E}_{0}$. Thermodynamically this term can be seen as the infinite
volume energy and field theoretically this corresponds to the sum of all
infrared substractions, which achieve the convergence of the sums (\ref{gb})
for large $r$.

\noindent \underline{$\mathbf{1/2<\Delta <1:}$} Now the individual integrals
in the expansion (\ref{O}) are still IR convergent, but cease to be UV
convergent. Nonetheless, we may still employ similar arguments as in the
previous regime and find again for the behaviour in (\ref{gb}) that $%
\mathcal{E}_{0}^{^{\prime }}=0$ and $f_{n}(r)=r^{2n(1-\Delta )}$ \cite%
{ADETBA}. From (\ref{gb}) and (\ref{T1}) follows once more that we can
identify $\left\langle T_{\,\,\,\,\mu }^{\mu }\right\rangle |_{r=0}=-\pi
^{2}/3\mathcal{E}_{0}$. However, now the field theoretic interpretation of
this term changes. Since we require in this case UV conterterms to make the
individual integrals finite, the $\mathcal{E}_{0}$-term corresponds now to
the sum of these UV counterterms and all infrared substractions, which
achieve the convergence of the sums (\ref{gb}) for large $r$. Indeed, for
the concrete models studied below this becomes visible in a change of sign
in the transition from the regime $\Delta <1/2$ to $\Delta >1/2$.

\noindent \underline{$\mathbf{\Delta <0:}$} Now the individual integrals in
the expansion (\ref{gb}) are still UV convergent, but cease to be IR
convergent even on a cylinder. General arguments now yield for the behaviour
in (\ref{gb}) that $\mathcal{E}_{0}\neq 0$, $\mathcal{E}_{0}^{^{\prime
}}\neq 0$ and $f_{n}(r)=(\alpha +\ln (r))^{-n}$ with $\alpha $ being some
constant \cite{Res,Marcio,FKS2,Fat1,Fat2}. One still finds that $%
\left\langle T_{\,\,\,\,\mu }^{\mu }\right\rangle |_{r=0}=-\pi ^{2}/3%
\mathcal{E}_{0}$, e.g.~\cite{Destri:1991ps}, but now the interpretation is
less obvious as some counterterms also accumulate in the $\mathcal{E}%
_{0}^{^{\prime }}$-term.

\noindent \underline{$\mathbf{\Delta =1/2:}$} In this case one usually finds
free Fermions in the model and $\mathcal{E}_{0}\neq 0$, $\mathcal{E}%
_{0}^{^{\prime }}\neq 0$, $f_{n}(r)=$ $r^{n}$. Now the vacuuum energy is
divergent, see e.g.~\cite{Klassen:1991dx} for an analytical expression.

We will investigate some concrete theories.

\section{$0<\Delta <1/2$, minimal affine Toda field theories}

These theories have been studied before \cite{Klassen:1991dx,Fateev:1994av},
nonetheless, we recall them here as they are easy examples which illustrate
the working of the above formulae and we shall also point out some novel
features. We recall first that minimal affine Toda field theories can be
realized as perturbations of the coset conformal field theories $\mathbf{g}%
_{1}\otimes \mathbf{g}_{1}/\mathbf{g}_{2}$, with $\mathbf{g}_{k}$\textbf{\ }%
being a simply laced Kac-Moody algebra of rank $\ell $ and level $k$ \cite%
{EY,HM}. The corresponding Virasoro central charges $c$ and conformal
dimension of the perturbing operator $\Delta $ are%
\begin{equation}
c=\frac{2\ell }{2+h}\qquad \quad \text{and\qquad \quad }\Delta =\frac{2}{2+h}%
~,
\end{equation}%
respectively. Apart from $h=2$, i.e.~the free Fermion with $\mathbf{g}=A_{1}$%
, we always have for the Coxeter number $h>2$ and are therefore in the
stated regime $0<\Delta <1/2$. The renormalization issues are handled most
easily in this case and the vacuum energies are computable with the above
arguments. With regard to assumption i), we recall the expansion of the
TBA-kernel for these theories \cite{Klassen:1991dx,Niedermeier,Fring:1996ge}%
\begin{equation}
\varphi _{ij}(\theta )=-4\sum\limits_{s\in \mathcal{E}}\cot \frac{s\pi }{h}%
x_{i}(s)x_{j}(s)e^{-s\left\vert \theta \right\vert }~,
\end{equation}%
with $\mathcal{E}$ =$\{s+nh\}$, $s$ being an exponent of $\mathbf{g}$, $n\in 
\mathbb{N}_{0}$ and $x_{i}(s)$ are the left eigenvectors of the Cartan
matrix. In particular, we have $x_{i}(1)=m_{i}/m$, with $m$ being an overall
mass scale, which is needed for the assumption ii) to hold. Having therefore
the quantity $\varphi _{ij}^{(1)}$ in the form (\ref{A3}), we deduce
immediately with the help of (\ref{T}) 
\begin{equation}
\left\langle T_{\,\,\,\,\mu }^{\mu }\right\rangle =m^{2}\frac{\pi }{2}\tan 
\frac{\pi }{h}~.  \label{Tmin}
\end{equation}%
Obviously apart from the free Fermion with $h=2$, when $\left\langle
T_{\,\,\,\,\mu }^{\mu }\right\rangle \rightarrow \infty $, we have $%
\left\langle T_{\,\,\,\,\mu }^{\mu }\right\rangle >0$. This result agrees
with a similar formula obtained in \cite{Klassen:1991dx} in terms of the
coefficients $\varphi _{11}^{(1)}$ without explicit evaluation and
\textquotedblleft $1$\textquotedblright\ referring to the lightest particle.
More concret case-by-case studies were carried out in \cite{Fateev:1994av}
for perturbations of $\mathbf{g}_{l}\otimes \mathbf{g}_{k}/\mathbf{g}_{k+l}$%
-coset CFT's (see formulae (3.14) therein). When using the overall mass
scale to perform suitable normalizations the formula for $k=l=1$ in there
can be brought into the universal formula (\ref{Tmin}), which is not obvious
at first sight. The formulae in \cite{Fateev:1994av} are expressed in terms
of a mass scale $M$ whose relation with our $m$ varies for every theory as 
\begin{equation}
\begin{array}{ll}
A_{\ell }: & M=m\sin \frac{\pi }{\ell -1} \\ 
D_{\ell }: & M=m/\sqrt{2} \\ 
E_{6}: & M=m\sqrt{\sqrt{\frac{3}{2}}\sin \frac{\pi }{12}} \\ 
E_{7}: & M=m\sqrt{\sin \frac{\pi }{18}/\sin \frac{2\pi }{9}} \\ 
E_{8}: & M=m\sqrt{2\sin \frac{\pi }{30}\sin \frac{\pi }{5}}~.%
\end{array}%
\end{equation}%
The advantage of our formulation relies on the fact that the masses are
normalized with respect to the same general mass scale $m$ for all simply
laced Lie algebras, which allows for the very compact and generic expression
(\ref{Tmin}). Alternatively these results were also confirmed in \cite{CKS}.

\section{$0<\Delta <1$, \textbf{g}$_{k}$-homogeneous sine-Gordon models}

Let us now consider a theory which is more interesting with regard to the
above mentioned problematic, namely the $\mathbf{g}_{k}$-HSG model \cite%
{Park:1994bx,Fernandez-Pousa:1997hi}, with $\mathbf{g}$\textbf{\ }being a
simple Lie algebra of rank $\ell $ and level $k$. These models can be viewed
as perturbed Wess-Zumino-Novikov-Witten (WZNW) \cite{Witten:1984ar}
coset-models 
\begin{equation}
\mathcal{S}_{\text{HSG}}=\mathcal{S}_{\text{WZNW}}+\frac{m^{2}}{\pi \beta
^{2}}\,\int d^{2}x\,\,\left\langle \Lambda _{+},g(\vec{x})^{-1}\Lambda _{-}g(%
\vec{x})\right\rangle \ .  \label{HSGaction}
\end{equation}%
Here $\left\langle \,\,,\,\,\right\rangle $ denotes the Killing form of $%
\mathbf{g}$ and $g(\vec{x})$ a group valued bosonic scalar field. $\Lambda
_{\pm }$ are arbitrary semi-simple elements of the Cartan subalgebra
associated with the maximal abelian torus $\mathbf{h}\subset \mathbf{g}$,
which have to be chosen not to be orthogonal to any root of $\mathbf{g}$.
The parameters $m$ and $\beta $ are the bare mass scale and the coupling
constant, respectively. The Virasoro central charge of the coset model and
the dimension of the perturbing operator are computed to 
\begin{equation}
c=\ell \,\frac{k\,h-h^{\vee }}{k+h^{\vee }}\qquad \quad \text{and\qquad
\quad }\Delta =\frac{h^{\vee }}{k+h^{\vee }}\;,  \label{cdata}
\end{equation}%
with $(h^{\vee })\,h$ being the (dual) Coxeter number of $\mathbf{g}$. We
note that now the constraint $\Delta <1/2$ does not automatically hold for
each level and the above mentioned complications could arise for some
theories in this series when changing from $k>h^{\vee }$ to $k<h^{\vee }$.
Up to now no indication for a different behaviour of the theories in this
two different regimes have been found in the literature. We treat the simply
laced and non-simply laced cases separately.

\subsection{Simply laced HSG-models}

As in the original formulation of these models, the algebra $\mathbf{g}$%
\textbf{\ }is assumed to be simply laced. Since for this case the expansion
of the kernel $\varphi $ does not appear in the literature, we will start
with the scattering matrix, which was found originally in \cite%
{Miramontes:1999hx} (see \cite{Castro-Alvaredo:1999em} for an integral
representation). We cast the matrix describing the scattering between the
particle of type $A=(a,\tilde{a})$ and $B=(b,\tilde{b})$, with $1\leq \tilde{%
a},\tilde{b}\leq \ell $; $1\leq a,b<k$ into the form 
\begin{equation}
S_{ab}^{\tilde{a}\tilde{b}}(\theta )=\eta _{ab}^{\tilde{a}\tilde{b}}\exp
\int \frac{dt}{t}\tilde{K}_{\tilde{a}\tilde{b}}(t)\frac{\sinh (at/k)\sinh
[(k-b)t/k]}{\sinh (t/k)\sinh t}e^{-it(\theta +\sigma _{\tilde{a}\tilde{b}%
})/\pi }~.  \label{SHSG}
\end{equation}%
Here $\eta _{ab}^{\tilde{a}\tilde{b}}=\exp [i\pi \varepsilon _{\tilde{a}%
\tilde{b}}(2-I_{A_{k-1}})_{\bar{a}b}^{-1}]$ are constant phase factors not
relevant for our analysis, $\tilde{K}_{\tilde{a}\tilde{b}}(t)=2\delta _{%
\tilde{a}\tilde{b}}\cosh t/k-I_{\tilde{a}\tilde{b}}$ with $I$ being the
incidence matrix of $\mathbf{g}$\textbf{\ }and the $\sigma $'s are the
resonance parameters, which indicate the presence of unstable particles in
these models. In order to evaluate the expansion for $\varphi $, we can
employ the residue theorem for a contour along the real axis closing up in
the positive half of the complex plane encircling all poles on the imaginary
axis in the upper half plane. Noting that in (\ref{SHSG}) $t=i\pi n$ are
simple poles, except for $t=i\pi nk$ which constitute double poles for $n\in 
\mathbb{N}$, we deduce for $\sigma _{\tilde{a}\tilde{b}}=0$ 
\begin{eqnarray}
\varphi _{ab}^{\tilde{a}\tilde{b}}(\theta ) &=&2\pi i\sum\limits_{s=1;s\neq
nk}^{\infty }\limfunc{Res}_{t=i\pi s}\left( -\frac{1}{\pi }\right) \tilde{K}%
_{\tilde{a}\tilde{b}}(t)\frac{\sinh (at/k)\sinh [(k-b)t/k]}{\sinh (t/k)\sinh
t}e^{-it\theta /\pi } \\
&=&-2\sum\limits_{s=1;s\neq nk}^{\infty }\tilde{K}_{\tilde{a}\tilde{b}}(i\pi
s)\frac{\sin (a\pi s/k)\sin (b\pi s/k)}{\sin (\pi s/k)}e^{-s\left\vert
\theta \right\vert }~.
\end{eqnarray}%
The desired coefficient (\ref{A3}) follows from this directly to%
\begin{equation}
\varphi _{\tilde{a}\tilde{b}}^{(1)}=2\tilde{K}_{\tilde{a}\tilde{b}}(i\pi
)m_{a}m_{b}/m^{2}\sin (\pi /k)
\end{equation}%
where $m_{a}^{\tilde{a}}=$ $m_{a}$ $m^{\tilde{a}}$ with $m_{a}=\sin a\pi /k$
being the masses of $A_{k-1}$-affine Toda field theory and $m^{\tilde{a}}$
are $\ell $ free mass scales. We choose them here to be all equal $m^{\tilde{%
a}}=m$ $\forall ~\tilde{a}$. Finally we derive from this a closed expression
for the vacuum expectation value for the trace of energy-momentum tensor%
\begin{equation}
\left\langle T_{\,\,\,\,\mu }^{\mu }\right\rangle =\pi m^{2}\sin (\pi
/k)\sum_{\tilde{a},\tilde{b} = 1}^{\ell }\left[ \tilde{K}^{-1}(i\pi )\right] _{%
\tilde{a}\tilde{b}}~.  \label{VACT}
\end{equation}%
We are not aware of a generic formulation for $\tilde{K}^{-1}(i\pi )$ and
analyse therefore the expression (\ref{VACT}) below in more detail
case-by-case. We can summarize our findings as%
\begin{equation}
\left\langle T_{\,\,\,\,\mu }^{\mu }\right\rangle \left\{ 
\begin{array}{l}
>0~~\qquad \text{for }k>h~~\equiv ~\Delta <1/2 \\ 
\rightarrow \infty \qquad \text{for }k=h~~\equiv ~\Delta =1/2 \\ 
<0~\qquad ~\text{for }k<h~\equiv ~\Delta >1/2%
\end{array}%
\right. ~.  \label{pos}
\end{equation}%
In many cases we can attribute the divergence for $\Delta =1/2$ to the
presence of free Ferminons. The change of sign when going from $\Delta <1/2$
to $\Delta >1/2$ reflects the fact that besides the IR counterterms, which
achieve the convergence of the sums (\ref{O}) for large $r$, needed in both
cases in the latter we also require UV conterterms to make the individual
integrals in (\ref{O}) finite.

\subsubsection{(A$_{\ell }$)$_{k}$-HSG model}

For A$_{\ell }$ the Coxeter number is $h=\ell +1$. The inverse of the matrix
relevant in (\ref{VACT}) can be cast in this case into a simple formula 
\begin{equation}
\left[ \tilde{K}_{A_{\ell }}^{-1}(i\pi )\right] _{\tilde{a}\tilde{b}}=\frac{%
\sin (\tilde{a}\pi /k)\sin [(h-\tilde{b})\pi /k]}{\sin (\pi /k)\sin (h\pi /k)%
}~\qquad \text{for }\tilde{a}\leq \tilde{b}.
\end{equation}%
Computing the sums over both entries then yields after some algebra%
\begin{equation}
\left\langle T_{\,\,\,\mu }^{\mu }\right\rangle =\frac{\pi m^{2}}{2\tan
^{2}\pi /2k}\left[ \tan \frac{h\pi }{2k}-h\tan \frac{\pi }{2k}\right] ~.
\label{vaca}
\end{equation}%
Hence, the condition $\left\langle T_{\,\,\,\mu }^{\mu }\right\rangle >0$
becomes 
\begin{equation}
\tan \frac{h\pi }{2k}>h\tan \frac{\pi }{2k}
\end{equation}%
or equivalently, when expanding the $\tan $,%
\begin{equation}
\frac{4}{\pi }\frac{h}{k}\sum\limits_{n=1}^{\infty }\frac{1}{%
(2n-1)^{2}-(h/k)^{2}}>h\frac{4}{\pi }\frac{1}{k}\sum\limits_{n=1}^{\infty }%
\frac{1}{(2n-1)^{2}-(1/k)^{2}}~.  \label{exp}
\end{equation}%
It is easily seen that (\ref{exp}) holds term by term once $h/k<1$, hence
establishing the first inequality in (\ref{pos}). Similar arguments show
that the opposite inequality holds in the regime $h/k>1$. We comment more on
the case $k=h$ below.

\subsubsection{(D$_{\ell }$)$_{k}$-HSG model}

For D$_{\ell }$ the Coxeter number is $h=2\ell -2$ and by evaluating (\ref%
{VACT}) similarly as in the previous subsection, we find 
\begin{equation}
\left\langle T_{\,\,\,\mu }^{\mu }\right\rangle =\frac{\pi m^{2}\sin \frac{%
\pi }{k}\left[ 2-(2+h)\cos \frac{h\pi }{2k}\right] +2\pi m^{2}\sin \frac{%
h\pi }{2k}}{\sin ^{2}\frac{\pi }{2k}\cos \frac{h\pi }{2k}}.  \label{vacd}
\end{equation}%
The condition $\left\langle T_{\,\,\,\mu }^{\mu }\right\rangle >0$ is now
equivalent to 
\begin{equation}
\sin \frac{\pi }{k}\left[ (2+h)-\frac{2}{\cos \frac{h\pi }{2k}}\right]
<2\tan \frac{h\pi }{2k}.  \label{cond}
\end{equation}%
Expanding the left and right hand side of this inequality yields by similar
arguments as in the previous subsection once more the relation (\ref{pos}).

\subsubsection{(E$_{6}$)$_{k}$-HSG model}

For E$_{6}$ the Coxeter number is $h=12$ and we find%
\begin{equation}
\left\langle T_{\,\,\,\mu }^{\mu }\right\rangle =2\pi m^{2}\frac{%
\sum\nolimits_{p=1}^{4}\tau _{p}\sin p\pi /k}{2\cos 4\pi /k-1}\qquad \quad 
\vec{\tau}~=(4,4,5,3).
\end{equation}%
We see that the numerator is $<0$ for $k=2$ and $>0$ for $k>2$. The
denominator is $>0$ for $k=2$, $k>12$ and $<0$ for $2<k<12$. The denominator
vanishes for $k=12$. Hence the relation (\ref{pos}) holds.

\subsubsection{(E$_{7}$)$_{k}$-HSG model}

For E$_{7}$ the Coxeter number is $h=18$ and we find%
\begin{equation}
\left\langle T_{\,\,\,\mu }^{\mu }\right\rangle =\frac{\pi
m^{2}\sum\nolimits_{p=1}^{7}\tau _{p}\sin p\pi /k}{\cos \pi /k(4\cos 6\pi
/k-2)}\qquad \quad \vec{\tau}~=(9,18,20,22,17,12,7)~.
\end{equation}%
We observe now that the numerator is $<0$ for $k<4$ and $>0$ otherwise. The
denominator on the other hand is $>0$ for $k=3,k>18$ and $<0$ otherwise
except for $k=2,18$ in which case it is zero. Hence (\ref{pos}) holds also
in this case.

\subsubsection{(E$_{8}$)$_{k}$-HSG model}

For E$_{8}$ the Coxeter number is $h=30$ and we find%
\begin{equation}
\left\langle T_{\,\,\,\mu }^{\mu }\right\rangle =\frac{\pi
m^{2}\sum\nolimits_{p=1}^{7}\tau _{p}\sin p\pi /k}{\cos 8\pi /k+\cos 6\pi
/k-\cos 2\pi /k-1/2}\qquad ~\vec{\tau}~=(4,8,12,12,13,10,7,4)~.
\end{equation}%
We see that the numerator is $<0$ for $k<5$ and $>0$ otherwise. The
denominator is $>0$ for $k=2,3,4;k>30$ and $<0$ otherwise except for $k=30$
in which case it is zero. Hence (\ref{pos}) holds also in this case.

\subsection{Non-simply laced HSG-models}

Now we allow the algebra $\mathbf{g}$\textbf{\ }to be also non-simply laced.
In this case the scattering matrix is slightly more complicated as the
symmetry between the exchange of long and short roots is lost. It can be
restored by the use of the symmetrizers $t_{\tilde{a}}$ of the incidence
matrix of $\mathbf{g}$, i.e.~$t_{\tilde{a}}I_{\tilde{a}\tilde{b}}=t_{\tilde{b%
}}I_{\tilde{b}\tilde{a}}$, with $t_{\tilde{a}}=2/\alpha _{\tilde{a}}^{2}$
and the length of long roots normalized to $\alpha _{l}^{2}=2$. In \cite%
{Korff:2000zu} the scattering between the particle of type $A=(a,\tilde{a})$
and $B=(b,\tilde{b})$, with $1\leq \tilde{a},\tilde{b}\leq \ell $; $1\leq
a,b<k_{\tilde{a}}=t_{\tilde{a}}k$ was proposed to be described by 
\begin{equation}
S_{ab}^{\tilde{a}\tilde{b}}(\theta )=\eta _{ab}^{\tilde{a}\tilde{b}}\exp
\int \frac{dt}{t}\tilde{K}_{\tilde{a}\tilde{b}}(t)\frac{\sinh (at/k_{\tilde{a%
}})\sinh [(1-b/k_{\tilde{b}})t]}{\sinh (t/k_{\tilde{a}\tilde{b}})\sinh t}%
e^{-it(\theta +\sigma _{\tilde{a}\tilde{b}})/\pi }~.  \label{Snon}
\end{equation}%
Here $\eta _{ab}^{\tilde{a}\tilde{b}}$ are once more constant phase factors
not relevant for our analysis. Furthermore, one needs the quantity $k_{%
\tilde{a}\tilde{b}}=k\max (t_{\tilde{a},}t_{\tilde{b}})$ and the matrix $%
\tilde{K}$ with entries $\tilde{K}_{\tilde{a}\tilde{b}}(t)=2\delta _{\tilde{a%
}\tilde{b}}\cosh t/k_{\tilde{a}}-I_{\tilde{a}\tilde{b}}t_{\tilde{b}}/\max
(t_{\tilde{a},}t_{\tilde{b}})$. Note that in comparison with \cite%
{Korff:2000zu} we interchanged the long and short roots, that is we have
taken the $t$'$s$ to be left and not the right symmetrizers of the incidence
matrix. A similar analysis as in the previous subsection leads now to the
following expansion of the TBA-kernel 
\begin{equation}
\varphi _{ab}^{\tilde{a}\tilde{b}}(\theta )=-2\sum\limits_{s=1;s\neq nk_{%
\tilde{a}\tilde{b}}}^{\infty }\tilde{K}_{\tilde{a}\tilde{b}}(i\pi s)\frac{%
\sin (a\pi s/k_{\tilde{a}})\sin (b\pi s/k_{\tilde{b}})}{\sin (\pi s/k_{%
\tilde{a}\tilde{b}})}e^{-s\left\vert \theta \right\vert }~,
\end{equation}%
such that%
\begin{equation}
\varphi _{\tilde{a}\tilde{b}}^{(1)}=2\tilde{K}_{\tilde{a}\tilde{b}}(i\pi
)m_{a}^{\tilde{a}}m_{b}^{\tilde{b}}/m^{2}\sin (\pi /k_{\tilde{a}\tilde{b}})~.
\end{equation}%
Here the masses are assumed to renormalize with an overall factor and are
therefore expected to be the same as in the semi-classical analysis \cite%
{Fernandez-Pousa:1998iu}, that is $m_{a}^{\tilde{a}}=m\sin a\pi /k_{\tilde{a}%
}$. The overall mass scales associated with each colour are once more
choosen to be the same. Thus we finally deduce%
\begin{equation}
\left\langle T_{\,\,\,\,\mu }^{\mu }\right\rangle =\pi m^{2}\sum_{\tilde{a},%
\tilde{b}=1}^{\ell }\left[ \hat{K}^{-1}\right] _{\tilde{a}\tilde{b}}~.
\label{Tnon}
\end{equation}%
where $\hat{K}_{\tilde{a}\tilde{b}}=\tilde{K}_{\tilde{a}\tilde{b}}(i\pi
)\sin (\pi /k_{\tilde{a}\tilde{b}})$. As in the previous case, we are not
aware of a generic formulation for $\hat{K}^{-1}$ and analyse therefore (\ref%
{Tnon}) in more detail case-by-case. Our findings are summarized as%
\begin{equation}
\left\langle T_{\,\,\,\,\mu }^{\mu }\right\rangle \left\{ 
\begin{array}{c}
>0\qquad ~\text{for }k>h^{\vee }~~\equiv ~\Delta <1/2 \\ 
\rightarrow \infty \qquad \text{for }k=h^{\vee }~\equiv ~\Delta =1/2 \\ 
<0\qquad ~\text{for }k<h^{\vee }~\equiv ~\Delta <1/2%
\end{array}%
\right. ~,  \label{Tm}
\end{equation}%
with similar interpretations as in (\ref{pos}). We establish (\ref{Tm}) in
more detail case-by-case.

\subsubsection{(G$_{2}$)$_{k}$-HSG model}

The dual Coxeter number for G$_{2}$ is $h^{\vee }=4$ and the symmetrizers
are taken to be $t_{1}=3,t_{2}=1$. With these data we compute from (\ref%
{Tnon}) 
\begin{equation}
\left\langle T_{\,\,\,\mu }^{\mu }\right\rangle =2\pi m^{2}\frac{\sin \pi
/k+\sin 4\pi /3k}{2\cos 4\pi /3k-1}~.
\end{equation}%
Obviously, the numerator is $>0$ for $k\geq 2$, whereas the denominator is $%
<0$ for $k=2,3$ and otherwise $>0$ except for $k=4$ when it is zero.
Evidently this agrees with (\ref{Tm}).

\subsubsection{(F$_{4}$)$_{k}$-HSG model}

The dual Coxeter number for F$_{4}$ is $h^{\vee }=9$ and the symmetrizers
are taken to be $t_{1}=t_{2}=1$ and $t_{3}=t_{4}=2$. From (\ref{Tnon}) we
compute 
\begin{equation}
\left\langle T_{\,\,\,\mu }^{\mu }\right\rangle =2\pi m^{2}\frac{%
2\sum\nolimits_{p=1}^{6}\sin p\pi /2k-\sin 3\pi /2k}{2\cos 3\pi /k-1}~.
\end{equation}%
The numerator is $>0$ for $k\geq 2$, whereas the denominator is $<0$ for $%
2\leq k<9$ and otherwise $>0$ except for $k=9$ when it is zero. Evidently
this agrees with (\ref{Tm}).

\subsubsection{(B$_{\ell }$)$_{k}$-HSG model}

The dual Coxeter number for B$_{\ell }$ is $h^{\vee }=2\ell -1$ and the
symmetrizers are taken to be $t_{1}=t_{2}=\ldots =t_{\ell -1}=2$ and $%
t_{\ell }=1$. We find now for even rank $\ell $%
\begin{equation}
\left\langle T_{\,\,\mu }^{\mu }\right\rangle =-\frac{\pi m^{2}}{\cos \frac{%
\pi }{2k}}\frac{\sum\limits_{p=2}^{h^{\vee }-1}\sin \frac{\pi p}{2k}+\frac{%
\ell }{2}\sin \frac{\pi h^{\vee }}{2k}+2\cos \frac{\pi }{2k}%
\sum\limits_{p=1}^{(\ell -2)/2}(\ell -2p-1)\sin \frac{\pi (1+h^{\vee }-4p)}{%
2k}}{1+2\sum_{p=1}^{\ell /2}(-1)^{p}\cos \frac{\pi p}{k}},
\end{equation}%
whereas for odd $\ell $ we obtain%
\begin{equation}
\quad \left\langle T_{\,\,\mu }^{\mu }\right\rangle =\frac{\pi m^{2}}{\cos 
\frac{\pi }{2k}}\frac{\sum\limits_{p=1}^{h^{\vee }-1}\sin \frac{\pi p}{2k}+%
\frac{\ell }{2}\sin \frac{\pi h^{\vee }}{2k}+2\cos \frac{\pi }{2k}%
\sum\limits_{p=1}^{(\ell -3)/2}(\ell -2p-1)\sin \frac{\pi (1+h^{\vee }-4p)}{%
2k}}{1+2\sum_{p=1}^{\ell /2}(-1)^{p}\cos \frac{\pi p}{k}}.
\end{equation}%
Once more we confirm (\ref{Tm}). As the details are rather involvolved we
drop them here.

\subsubsection{(C$_{\ell }$)$_{k}$-HSG model}

The dual Coxeter number for C$_{\ell }$ is $h^{\vee }=\ell +1$ and the
symmetrizers are taken to be $t_{1}=t_{2}=\ldots =t_{\ell -1}=1$ and $%
t_{\ell }=2$. 
\begin{eqnarray}
\left\langle T_{\,\,\mu }^{\mu }\right\rangle &=&\frac{\pi m^{2}(i)^{\ell }}{%
\cos \frac{\pi }{2k}}\left[ \frac{\sum_{p=1}^{h^{\vee }-1}(p-1)\sin \frac{%
\pi p}{2k}+\frac{\ell }{2}\sin \frac{\pi h^{\vee }}{2k}}{1+2\sum_{p=1}^{\ell
/2}(-1)^{p}\cos \frac{\pi p}{k}}\right] ,\quad \text{for \ }\ell \text{%
\thinspace\ even} \\
\left\langle T_{\,\,\mu }^{\mu }\right\rangle &=&\frac{\pi m^{2}}{\cos \frac{%
\pi h^{\vee }}{2k}}\left[ \sum_{p=2}^{h^{\vee }-1}(p-1)\sin \frac{\pi p}{2k}+%
\frac{\ell }{2}\sin \frac{\pi h^{\vee }}{2k}\right] ,\quad \text{for \ }\ell 
\text{\thinspace\ odd.}
\end{eqnarray}%
Once more we confirm (\ref{Tm}) and drop the details for the same reasons as
in the previous subsection.

\section{$\Delta =1/2$, $g_{h^{\vee }}$-homogeneous sine-Gordon model}

The case $\Delta =1/2$ is very special as then the vacuum energy diverges in
the extreme UV limit. Such type of behviour is well known from free Fermions
in form of logarithmic ultraviolet singularities, meaning that (\ref{T1})
yields $\left\langle T_{\,\,\,\mu }^{\mu }\right\rangle \rightarrow \infty $
for $r\rightarrow 0$. Explicit analytic formulae for the free Fermion $c(r)$%
-function can be found in \cite{Klassen:1991dx}. Indeed in many cases we can
make this connection quite explicit. It suffices to present an examples to
illustrate this point.

\subsection{$(A_{\ell })_{\ell +1}$-HSG theories}

Let us have a closer look at the $(A_{\ell })_{\ell +1}$-HSG theories in
order to see how the Fermions arise in there. Obviously for $h=k$ the
expression (\ref{vaca}) yields $\left\langle T_{\,\,\,\mu }^{\mu
}\right\rangle \rightarrow \infty $. Already in \cite{Castro-Alvaredo:1999em}
it was noticed that the $(A_{2})_{3}$-HSG model decomposes into four free
Fermions when the resonance parameter vanishes. From the fact that the
central charge (\ref{cdata}) becomes in general $\ell ^{2}/2$ for $(A_{\ell
})_{\ell +1}$-HSG models, one might suspect that they always decompose
completely into $\ell ^{2}$ free Fermions for vanishing resonance
parameters, such that each Fermion contributes $1/2$ to the central charge.
However, this is not quite the case as the following argument shows.

In order to count the Fermions, identified here simply with the amount of
particles which contribute $1/2$ to the central charge, we recall the
constant TBA equations, which arise from (\ref{TBA}) after some standard
manipulations. For the $(A_{\ell })_{\ell +1}$-HSG models they take on the
form 
\begin{equation}
x_{a}^{\tilde{a}}=\prod\limits_{b,\tilde{b}=1}^{\ell }(1+x_{b}^{\tilde{b}%
})^{N_{ab}^{\tilde{a}\tilde{b}}}\,\,\text{\qquad with\quad\ }N_{ab}^{\tilde{a%
}\tilde{b}}=\delta _{ab}\delta _{\tilde{a}\tilde{b}}-\left( K_{A_{\ell
}}^{-1}\right) _{\tilde{a}\tilde{b}}\left( K_{A_{\ell }}\right) _{ab}~.
\label{cTBA}
\end{equation}

\noindent Solving these equations for the $x_{a}^{\tilde{a}}$ =$\exp
(-\varepsilon _{a}^{\tilde{a}})$ yields the effective central charge as 
\begin{equation}
c_{\text{eff}}=\frac{6}{\pi ^{2}}\sum\limits_{a,\tilde{a}=1}^{\ell }\mathcal{%
L}\left( \frac{x_{a}^{\tilde{a}}}{1+x_{a}^{\tilde{a}}}\right) =\frac{\ell
^{2}}{2}  \label{ceff}
\end{equation}%
with $\mathcal{L}(x)=\sum_{n=1}^{\infty }x^{n}/n^{2}+\ln x\ln (1-x)/2$
denoting Rogers dilogarithm. The solutions of (\ref{cTBA}) are very simple
in this case%
\begin{equation}
x_{a}^{\tilde{a}}=\frac{\sin [\pi \tilde{a}/(1+\ell )]}{\sin [\pi a/(1+\ell
)]}~.  \label{xaa}
\end{equation}%
Therefore we have $x_{a}^{a}=x_{a}^{\ell +1-a}=1$ and since $\mathcal{L}%
(1/2)=\pi ^{2}/12$ it follows from this that each of the particles $%
(a,a),(a,\ell +1-a)$ for $1\leq a\leq \ell $ contributes $1/2$ to the
effective central charge in (\ref{ceff}). Hence in the $(A_{\ell })_{\ell +1}
$-HSG models we have always $2\ell $ or $2\ell -1$ free Fermions when $\ell $
is odd or even, respectively. The remaining particles can be organized
without exceptions in pairs $(a,\tilde{a}),(\tilde{a},a)$. Noting with (\ref%
{xaa}) that obviously $x_{a}^{\tilde{a}}=(x_{\tilde{a}}^{a})^{-1}$ and
recalling the fact that $\mathcal{L}(x)+\mathcal{L}(1-x)=\pi ^{2}/6$
explains then that the central charge has to be an integer or a semi-integer
for these models.

In general, it is less straightforward for the other algebras to identify
particles which directly contribute $1/2$ to the central charge. In fact,
mostly the particles occur in pairs, triplets or higher multiplets
contributing integers or semi-integer values to $c$. 

\section{$\Delta <0$, affine Toda field theories}

Affine Toda field theories related to simply laced and non-simply laced Lie
algebras have a quite different behaviour due to the fact that in the first
case all masses renormalize with an overall factor, which is not the case in
the latter (see e.g.~\cite{Braden:1990bu}). As a result of this, the
strong-weak duality observed for ATFT related to simply laced algebras is
broken for those associated with non-simply laced Lie algebras. Despite the
fact that there exists a uniform formulation, we will treat them here
separately as this will be more transparent.

\subsection{ Simply laced}

ATFT are quite well studied examples of integrable models, which can be
viewed in the spirit of (\ref{H}) which was noted first in \cite{HM,EY} 
\begin{equation}
\mathcal{S}_{ATFT}\mathcal{=}\int d^{2}x\frac{1}{2}(\partial _{\mu }\vec{%
\varphi})^{2}+\mu \sum_{i=0}^{\ell }n_{i}e^{\beta ~\vec{\alpha}_{i}\cdot 
\vec{\varphi}}~.  \label{SAT}
\end{equation}%
The fixed point part of the action $\mathcal{S}_{CFT}$ corresponds to the
conformal Toda field theories when the sum over the simple roots $\vec{\alpha%
}_{i}$ starts at $i=1$. The $\mu ,\beta $ are real parameters and the $n_{i}$
are the Kac labels related to the negative of the highest root $\vec{\alpha}%
_{0}=-\sum\nolimits_{i=1}^{\ell }n_{i}\vec{\alpha}_{i}$. The Virasoro
central charge of the conformal Toda field theories and the conformal
dimension of the perturbing operator $V=\mu n_{0}e^{\beta ~\vec{\alpha}%
_{0}\cdot \vec{\varphi}}$ have been computed in \cite{HM}

\begin{equation}
c=\ell +\frac{4\ell h(h+1)}{B(2-B)}\qquad \quad \text{and\qquad \quad }%
\Delta =1-\frac{2h}{2-B}~,  \label{ko}
\end{equation}%
where we use the effective coupling\footnote{%
Confusion arises sometimes due to different conventions. For instance we can
relate our notations to the ones used in \cite{Fateev1} by a simple
rescaling of the fields $\varphi =\varphi _{F}/\sqrt{4\pi }$ compensated by
a scaling of the coupling constant $\beta =b_{F}/\sqrt{4\pi }$. In addition,
one takes the effective coupling constant to be $B=2B_{F}$.} $0\leq B=2\beta
^{2}/(\beta ^{2}+4\pi )\leq 2$. Since $2h>1-B/2$ is always true we are in
the regime $\Delta <0$ and expect the above mentioned complications with
regard to renormalization to arise. To establish that, we recall first \cite%
{Klassen:1991dx,Niedermeier,Fring:1996ge}

\begin{equation}
\varphi _{ij}(\theta )=-2\sum\limits_{s\in \mathcal{E}}\sin \frac{s\pi B}{2h}%
\sin \frac{s\pi (2-B)}{2h}/\sin \frac{s\pi }{h}x_{i}(s)x_{j}(s)e^{-s\left%
\vert \theta \right\vert }~,
\end{equation}%
and deduce thereafter from (\ref{A3}) and (\ref{T}) 
\begin{equation}
\left\langle T_{\,\,\,\,\mu }^{\mu }\right\rangle =\frac{\pi m^{2}\sin (\pi
/h)}{\sin (\pi B/2h)\sin [\pi (2-B)/2h]}~.  \label{tg}
\end{equation}%
Clearly, as $0\leq B\leq 2$ we have $\left\langle T_{\,\,\,\,\mu }^{\mu
}\right\rangle >0$. Up to an overall mass re-scaling of $m\rightarrow 2m$,
this agrees precisely with the results in \cite{Destri:1991ps}, which were
obtained by matching the high-energy behaviour of the scattering matrix with
a Feynman diagramatic analysis.

It is very interesting to note that by an analytic continuation from real to
purely complex coupling we can also reach the regime for $\Delta >0$ for (%
\ref{tg}) and observe similar phenomena as for the HSG-models\footnote{%
We are grateful to Al.~B. Zamolodchikov for pointing this out to us.}. For $%
h=2$ this means we continue from sinh-Gordon to sine-Gordon. Following for
this case the argumentation of Destri and De Vega \cite{Destri:1991ps}, we
relate the sinh-Gordon coupling $\beta $ to the sine-Gordon coupling $\tilde{%
\beta}$ via $\beta \rightarrow i\tilde{\beta}/\sqrt{2}$ according to the
standard conventions. Also we replace the breather mass scale $m$ with the
soliton mass scale $\tilde{m}$ via $m^{2}\rightarrow 4\tilde{m}^{2}\sin
^{2}\pi B/2$ such that we end up with the simple formula%
\begin{equation}
\left\langle T_{\,\,\,\,\mu }^{\mu }\right\rangle =\pi \tilde{m}^{2}\tan 
\frac{\pi }{2}\left( \frac{\Delta }{\Delta -1}\right) ~,  \label{TSG}
\end{equation}%
where $\Delta =$ $\tilde{\beta}^{2}/8\pi $ is the conformal dimension of the
perturbing $\cos $-term in the model. This agrees also with \cite%
{Zamolodchikov:1995xk}. Note that in the previous argument we
considered sinh-Gordon as a perturbed Liouville theory, whereas now we
perturb the free theory rather than complex Liouville. Analysing (\ref{TSG})
in more detail one observes%
\begin{equation}
\left\langle T_{\,\,\,\,\mu }^{\mu }\right\rangle \left\{ 
\begin{array}{rrr}
<0 & \qquad \text{for} & \quad \frac{2n-2}{2n-1}<\Delta <\frac{2n-1}{2n} \\ 
\rightarrow \infty & \text{for} & \Delta =\frac{2n-1}{2n} \\ 
>0 & \text{for} & \frac{2n-1}{2n}<\Delta <\frac{2n}{2n+1} \\ 
=0 & \text{for} & \Delta =\frac{2n}{2n+1}%
\end{array}%
\right.
\end{equation}%
with $n\in \mathbb{N}$. Note that in particular for $n=1$ we have as for the
homogeneous sine-Gordon model at $\Delta =1/2$ a transition point at which
the sign changes by passing through a singularity. Moreover, precisely this
value corresponds to the free Fermion point, which in this case is a very
explicit example for the free Fermion picture advocated above. However, in
this case the structure is more complicated as first of all we have an
infinite number of such points rather than just one as in the HSG-models. In
addition $\left\langle T_{\,\,\,\,\mu }^{\mu }\right\rangle $ is not always
divergent at these points, but can also vanish.

\subsection{Non-simply laced}

It is known, that the above mentioned complication of mass renormalization
is reconciled if one views ATFT's in terms of dual pairs of Lie algebras.
Since simply laced Lie algebras are self-dual, this picture does not yield
anything new for that case. The dual pairs of non-simply laced Lie algebras
are $(G_{2}^{(1)},D_{4}^{(3)})$, $(F_{4}^{(1)},E_{6}^{(2)})$, $(B_{\ell
}^{(1)},A_{2\ell -1}^{(2)})$ and $(C_{\ell }^{(1)},D_{\ell +1}^{(2)})$. Each
algebra of these pairs allows for a description of the form (\ref{SAT})
related to each other by the strong-weak duality transformation $\beta
\rightarrow 4\pi /\beta $, where the untwisted algebras relate to the weak
coupling limit. The vacuum energies associated to all non-simply laced
affine Toda theories were stated in \cite{Fat2}. As in there no details were
presented on how they were obtained, it will be instructive to show that the
procedure outlined in section 2, also holds in this case.

Let us first of all see what we have to expect with regard to the arguments
outlined above and compute the Virasoro central charge and the dimension of
the perturbing operator. According to \cite{Fat2} we have 
\begin{equation}
c=\ell +12\vec{Q}^{2}\quad \quad \text{with\quad \quad }\vec{Q}=\beta \vec{%
\rho}+\frac{1}{\beta }\vec{\rho}^{\vee },  \label{CN}
\end{equation}%
with ($\vec{\rho}^{\vee }$)\textbf{\ }$\vec{\rho}$\textbf{\ }being the
(dual) Weyl vector of the untwisted Lie algebra given by half the sum of the
positive (co)roots. Note that when evaluating (\ref{CN}) for the simply
laced case yields precisely (\ref{ko}), but for the non-simply laced case it
differs from the expressions in \cite{HM} by the use of $\vec{\rho}^{\vee }$
rather than always $\vec{\rho}$. The confomal dimension of a spinless
primary field $V_{\vec{a}}(x)=e^{(\vec{Q}+\vec{a})\cdot \mathbf{\vec{\varphi}%
(}\vec{x}\mathbf{)}}$ in the underlying CFT is $\Delta (\vec{a})=(\vec{Q}%
^{2}-\vec{a}^{2})/2$, such that the perturbing field $\mu n_{0}V_{(\beta 
\vec{\alpha}_{0}-\vec{Q})}(x)$ has conformal dimension 
\begin{equation}
\Delta (\beta \vec{\alpha}_{0}-\vec{Q})=\beta \vec{\alpha}_{0}.\vec{Q}-\frac{%
\beta ^{2}\vec{\alpha}_{0}^{2}}{2},  \label{DN}
\end{equation}%
$\vec{\alpha}_{0}$ defined as in the previous section, that is being the
negative of the highest root. It will turn out that these dimension will
always be smaller zero. We will compute the precise values for some concrete
examples below.

Unlike to the previous cases the expansion for the kernel $\varphi $ does
not appear in the literature, we therefore start here with the scattering
matrix, which can be cast into the universal form \cite{Oota:1997un,FKS} 
\begin{equation}
S_{ij}(\pm \theta >0)=\exp \left[ \mp 8\int \frac{dt}{t}\sinh (\vartheta
_{h}t)\sinh (t_{j}\vartheta _{H}t)\left[ K^{-1}(t)\right] _{ij}\,e^{\pm
it\theta /\pi }\right] ,  \label{Sn}
\end{equation}%
where $\vartheta _{h}=(2-B)/2h$, $\vartheta _{H}=B/2H$ with $h$ being the
Coxeter number of the untwisted algebra and $H$ its dual Coxeter number $%
h^{\vee }$ multiplied by the twist of the second algebra. The effective
coupling is now generalized to $B=2H\beta ^{2}/(H\beta ^{2}+4\pi h)$. The $%
t_{i}$ are the symmetrizers of the incidence matrix of the untwisted algebra 
$t_{i}I_{ij}=t_{j}I_{ji}$, with $t_{i}=2/\vec{\alpha}_{i}^{2}$ and the
length of long roots normalized to $\vec{\alpha}_{l}^{2}=2$. Also needed in (%
\ref{Sn}) is the inverse of the $q$-deformed Cartan matrix $K_{ij}(t)=\left(
q\bar{q}^{t_{j}}+q^{-1}\bar{q}^{-t_{j}}\right) \delta _{ij}-\left[ I_{ij}%
\right] _{\bar{q}}$ with deformation parameters $q=\exp (t\vartheta _{h})$, $%
\bar{q}=\exp (t\vartheta _{H})$ and $\left[ I_{ij}\right] _{\bar{q}}$ $=(%
\bar{q}^{I_{ij}}-\bar{q}^{-I_{ij}})/(\bar{q}-\bar{q}^{-1})$.

In order to evaluate the expansion for $\varphi $, we can employ once again
the residue theorem for a contour along the real axis closing up in the
positive half of the complex plane encircling all poles on the imaginary
axis in the upper half plane. Recalling that $\det K(t)=\prod\nolimits_{s\in 
\mathcal{E}}4\cosh \left[ (t+i\pi s)/2h\right] \cosh \left[ (t-i\pi s)/2h%
\right] $, we know the positions of all poles and it follows from the
integral representation (\ref{Sn}) that the TBA kernels admit a series
expansion of the form 
\begin{equation}
\varphi _{ij}(\theta )=16\,i\sum_{s\in \mathcal{E}}\underset{t\rightarrow
i\pi s}{\limfunc{Res}}\left[ \sinh (\vartheta _{h}t)\sinh (t_{j}\vartheta
_{H}t\,)\check{K}(t)_{ij}\,/\det K(t)e^{it\theta /\pi }\right] ~.
\label{ker}
\end{equation}%
We do not have a closed formula for the cofactors $\check{K}$, but for the
sake of our argument it will be sufficient here to present some examples.

\subsubsection{$(G_{2}^{(1)},D_{4}^{(3)})$-ATFT}

Let us first compute (\ref{CN}) and (\ref{DN}). We carry out the
computations in terms of the quantities of the untwisted algebra $%
G_{2}^{(1)} $ for which we have two simple roots $\vec{\alpha}_{1}$ and%
\textbf{\ }$\vec{\alpha}_{2}$ normalised as $\vec{\alpha}_{2}^{2}=2=3\vec{%
\alpha}_{1}^{2}$. Furthermore, the Weyl vector, its dual and the negative of
the highest root are given by 
\begin{equation}
\vec{\rho}=5\vec{\alpha}_{1}+3\vec{\alpha}_{2}\text{ ,}\quad \text{\quad }%
\vec{\rho}^{\vee }=5\vec{\alpha}_{1}+\vec{\alpha}_{2}\text{\quad and\quad }%
\mathbb{\alpha }_{0}=-3\vec{\alpha}_{1}-2\vec{\alpha}_{2}~.
\end{equation}%
These realizations allow to compute the quantities needed in (\ref{CN}) and (%
\ref{DN}), that is $3\vec{\rho}^{2}=14,$ $3\vec{\rho}^{\vee }\mathbf{.}\vec{%
\rho}^{\vee }=26$ and $3\vec{\rho}\mathbf{.}\vec{\rho}^{\vee }=3\vec{\rho}%
^{\vee }\mathbf{.}\vec{\rho}=8$. It follows therefore 
\begin{equation}
c=2+32\left[ \frac{13+3B(B-3)}{B(2-B)}\right] \text{\quad and\quad }\Delta =%
\frac{3B+2}{B-2}.
\end{equation}%
Clearly for $0<B<2$ we have $-\infty \leq \Delta \leq -1$.

To proceed further we need the (generalized) Coxeter number for this theory,
which are $h=6$ and $H=12$. The symmetrizers are $t_{1}=3$ and $t_{2}=1$.
Evaluating (\ref{ker}) and reading off the first order coefficient we obtain 
\begin{equation}
\varphi _{ab}^{(1)}=-8\sqrt{3}\frac{\sin \frac{(2-B)\pi }{12}\sin \frac{B\pi 
}{8}}{\cos \frac{\pi }{6}(1-\frac{B}{4})}\frac{m_{a}m_{b}}{m^{2}}\qquad
a,b=1,2,  \label{ker1}
\end{equation}%
where we normalized the masses to 
\begin{equation}
m_{1}=m\cos \frac{\pi }{6}(1+\frac{B}{4})\qquad \text{and\qquad }m_{2}=m~.
\label{mass}
\end{equation}%
We deduce then with (\ref{T}) 
\begin{equation}
\left\langle T_{\,\,\mu }^{\mu }\right\rangle =\frac{\pi m^{2}\cos \frac{\pi 
}{6}(1-\frac{B}{4})}{4\sqrt{3}\sin \frac{(2-B)\pi }{12}\sin \frac{B\pi }{8}}.
\label{tmumu}
\end{equation}%
Agreement with the results in \cite{Fat2} is achieved by changing to the
conventions used in there. For this one needs to re-define the effective
coupling to $B\rightarrow B^{\prime }=3B/(4+B)$ and introduce a
\textquotedblleft floating\textquotedblright\ Coxeter number $H^{\prime
}=(1-B^{\prime })h+B^{\prime }h^{\vee }$.

\subsubsection{$(F_{4}^{(1)},E_{6}^{(2)})$-ATFT}

For $F_{4}^{(1)}$ we normalize the four simple roots to $\vec{\alpha}%
_{1}^{2}=\vec{\alpha}_{2}^{2}=2\vec{\alpha}_{3}^{2}$ $=2\vec{\alpha}%
_{4}^{2}=2$. The Weyl vector, its dual and the negative of the highest root
are in this case given by 
\begin{eqnarray}
\vec{\rho} &=&16\vec{\alpha}_{1}+30\vec{\alpha}_{2}+42\vec{\alpha}_{3}+22%
\vec{\alpha}_{4}\text{ ,}\quad \ \ \vec{\rho}^{\vee }=\vec{\rho}+22\vec{%
\alpha}_{4}\text{ ,}  \notag \\
\text{ and\quad\ \ }\vec{\alpha}_{0} &=&-2\vec{\alpha}_{1}-3\vec{\alpha}%
_{2}-4\vec{\alpha}_{3}-2\vec{\alpha}_{4}\text{ }
\end{eqnarray}%
such that $\vec{\rho}^{2}=39,$ $\vec{\rho}^{\vee }\mathbf{.}\vec{\rho}^{\vee
}=402$ and $\vec{\rho}\mathbf{.}\vec{\rho}^{\vee }=\vec{\rho}^{\vee }\mathbf{%
.}\vec{\rho}=55.$ With this we find 
\begin{equation}
c=4+12\left[ \frac{1608+B(331B-1388)}{(2-B)B}\right] \text{\quad and\quad }%
\Delta =\frac{16+B}{B-2}.
\end{equation}%
Therefore $-\infty \leq \Delta \leq -8$.

For this theory we have $h=12$, $H=18,$ $t_{1}=t_{2}=1$ and $t_{3}=t_{4}=2.$
The ratios between the masses of the four particles in the theory are 
\begin{equation}
\frac{m_{4}}{m_{1}}=2\sin \frac{\pi }{4}(1+\frac{B}{18}),\text{ \ }\frac{%
m_{3}}{m_{1}}=1+2\cos \frac{\pi }{6}(1-\frac{B}{6}),\text{ \ }\frac{m_{2}}{%
m_{1}}=2\cos \frac{\pi }{12}(1-\frac{B}{6}).  \label{massf}
\end{equation}%
We choose the normalization such that $m_{1}=m$ and obtain from (\ref{ker}) 
\begin{equation}
\varphi _{ab}^{(1)}=-8\sqrt{3}\frac{\sin \frac{(2-B)\pi }{24}\sin \frac{B\pi 
}{18}}{\cos \frac{\pi }{4}(1-\frac{B}{18})}\frac{m_{a}m_{b}}{m^{2}}\qquad
a,b=1,2,3,4.  \label{kerf}
\end{equation}%
Therefore with (\ref{T}) we get$894/4=\allowbreak \frac{447}{2}$ 
\begin{equation}
\left\langle T_{\,\,\mu }^{\mu }\right\rangle =\frac{\pi m^{2}\cos \frac{\pi 
}{4}(1-\frac{B}{18})}{4\sqrt{3}\sin \frac{(2-B)\pi }{24}\sin \frac{B\pi }{18}%
}.  \label{tfmm}
\end{equation}%
We can match with the formulae in \cite{Fat2} by $B\rightarrow B^{\prime
}=4B/(6+B)$, $\tilde{H}=3(4-B^{\prime })$, $m_{1}\rightarrow m_{1}^{\prime }$%
, $m_{2}\rightarrow m_{3}^{\prime }$, $m_{3}\rightarrow m_{4}^{\prime }$ and 
$m_{4}\rightarrow m_{2}^{\prime }$.

\subsubsection{$(B_{2}^{(1)},A_{3}^{(2)})$-ATFT}

Let us now present the simplest example of the family $(B_{\ell
}^{(1)},A_{2\ell -1}^{(2)})$. In general, we choose for the algebra $B_{\ell
}^{(1)}$ the normalizations $\vec{\alpha}_{i}^{2}=2$ for $i=1,\ldots ,\ell
-1 $ and \ $\vec{\alpha}_{\ell }^{2}=1$. Then we have 
\begin{equation}
2\vec{\rho}=3\vec{\alpha}_{1}+4\vec{\alpha}_{2}\text{, \ }\quad 2\vec{\rho}%
^{\vee }=3\vec{\alpha}_{1}+8\vec{\alpha}_{2}\mathbf{\quad \ }\ \text{and \
\quad }\vec{\alpha}_{0}=-\vec{\alpha}_{1}-2\vec{\alpha}_{2},
\end{equation}%
from which we compute $12\vec{\rho}^{2}=30,$ $\vec{\rho}^{\vee }\mathbf{.}%
\vec{\rho}^{\vee }=\vec{\rho}^{2}+72$ and $\vec{\rho}\mathbf{.}\vec{\rho}%
^{\vee }=\vec{\rho}^{2}+4$. Therefore 
\begin{equation*}
c=2+8\left[ \frac{447+24B(4B-17)}{B(2-B)}\right] \text{\quad and\quad }%
\Delta =\frac{B+4}{B-2}~.
\end{equation*}%
Hence $-\infty \leq \Delta \leq -2$

\bigskip\ For this theory we have $h=4,$ $H=6,$ $t_{1}=1$ and $t_{2}=2.$ The
masses satisfy 
\begin{equation}
\frac{m_{1}}{m_{2}}=2\sin \frac{\pi }{4}(1+\frac{B}{6}),
\end{equation}%
and we choose $m_{1}=m$. Evaluating (\ref{ker}) we obtain now 
\begin{equation}
\varphi _{ab}^{(1)}=-8\frac{\sin \frac{(2-B)\pi }{8}\sin \frac{B\pi }{6}}{%
\sin \frac{\pi }{4}(1+\frac{B}{4})}\frac{m_{a}m_{b}}{m^{2}}\qquad a,b=1,2,
\label{kerb}
\end{equation}%
and therefore with (\ref{T}) 
\begin{equation}
\left\langle T_{\,\,\mu }^{\mu }\right\rangle =\frac{\pi m^{2}\sin \frac{\pi 
}{4}(1+\frac{B}{4})}{4\sin \frac{(2-B)\pi }{8}\sin \frac{B\pi }{6}}.
\label{tbmm}
\end{equation}%
Defining once more $B\rightarrow B^{\prime }=4B/(6+B)$ and $H=4-B^{\prime }$
we find agreement with \cite{Fat2}. The previous results also hold for the $%
(C_{2}^{(1)},D_{3}^{(2)})$-theory by exchanging the roles of particles $1$
and $2$, since the Dynkin diagrams of $B_{2}^{(1)}$ and $C_{2}^{(1)}$ are
identical up to the exchange of the short and the long root.

These examples are sufficient to support the validity of the approach
outlined in section 2.

\section{Conclusions}

We used the thermodynamic Bethe ansatz to compute vacuum energies $%
\left\langle T_{\,\,\mu }^{\mu }\right\rangle $ for various types of
perturbed conformal field theories. Despite the fact, that the models
considered exhibit different general behaviours, the assumption i)-iii),
needed for the validity of the approximations in the TBA, hold in all cases.

The general behaviour of $\left\langle T_{\,\,\mu }^{\mu }\right\rangle $ is
shown to be sensitive to IR and UV-conterterms, whose presence can be
characterized by the conformal scaling dimension $\Delta $ of the perturbing
operator. In the regime $0<\Delta <1/2$, realized by minimal ATFT and $%
\mathbf{g}_{k}$-HSG models for $k>h^{\vee }$, the quantity $\left\langle
T_{\,\,\mu }^{\mu }\right\rangle $ can be identified with the
IR-counterterms needed to compensate the divergencies in the perturbative
series expansion (\ref{O}), when viewed on a cylinder. In contrast, in the
regime $1/2<\Delta <1$, realized by $\mathbf{g}_{k}$-HSG models for $%
k<h^{\vee }$, the quantity $\left\langle T_{\,\,\mu }^{\mu }\right\rangle $
can be associated to the sum of the aforementioned IR counterterms and UV
counterterms needed to guarantee the finiteness of the individual integrals
in the expansion. In the models studied here these additional counterterms,
when passing from $\Delta <1/2$ to $\Delta >1/2$ show up in a change of sign
in $\left\langle T_{\,\,\mu }^{\mu }\right\rangle $. It would be extremely
interesting to verify this assertion by some explicit perturbative
computations for the HSG-models. For the regime $\Delta <0$, realized here
by the ATFT (simply laced as well as non-simply laced) $\left\langle
T_{\,\,\mu }^{\mu }\right\rangle $ constitutes a mixture of several types of
counterterms, less obvious to disentangle. The divergence of $\left\langle
T_{\,\,\mu }^{\mu }\right\rangle $ at $\Delta =1/2$ can be attributed to the
occurrence of free Fermions, for which such type of behaviour is well known
from explicit analytical expressions. However, we were not able to identify
the free Fermions in all $\mathbf{g}_{h^{\vee }}$-HSG models, which can be
viewed as perturbed CFT's with $\Delta =1/2$. This needs further
investigations.

\textbf{Acknowledgments: }This work is supported by the EU network
\textquotedblleft EUCLID, \emph{Integrable models and applications: from
strings to condensed matter}\textquotedblright , HPRN-CT-2002-00325 and was
partly carried out at the Freie Universit\"{a}t Berlin where it was
supported by the Sonderforschungsbereich SFB288. We are grateful to Al.~B.
Zamolodchikov for very useful discussions.


\end{document}